# Emergence of Tsallis Statistics from a Self-Referential Nonlinear Operator: A Variational Framework

**Lucio Marassi**

*School of Science and Technology (ECT), Federal University of Rio Grande do Norte (UFRN), Natal, RN, Brazil*

lucio.marassi@ufrn.br

**Abstract**

We develop a variational thermodynamic framework for statistical systems governed by a self-referential nonlinear operator $\Omega$ characterized by structural exponents $\alpha > 0$, $\beta \geq 0$, a symmetric kernel K, and a self-coupling constant $\kappa \geq 0$. The central object is the self-consistency entropy S[Ψ] = −D_{KL}(Ψ ‖ ΩΨ), which vanishes at the fixed points of $\Omega$ and serves as a natural Lyapunov functional. Within the local kernel (mean-field) approximation, minimization of the free energy $F = U - TS$ admits the Tsallis q-exponential distribution as an equilibrium state, with the entropic index $q = \alpha + \beta$ emerging directly from the fixed-point structure of the operator rather than being postulated. The framework yields a consistent thermodynamic description, including a generalized equation of state $PV = (2-q)T$, response functions, and a critical temperature associated with spontaneous symmetry breaking. The relation $q = \alpha + \beta$ connects independently measurable structural exponents of the feedback mechanism to the observed tail index, providing a parameter-free criterion that distinguishes this approach from superstatistics, constrained entropy maximization, and q-deformed formalisms. This work establishes an operator-theoretic foundation for nonextensive statistical mechanics in which nonlinear self-referential feedback naturally generates Tsallis statistics in the mean-field limit.

## 1. Introduction

A defining feature of many complex physical and biological systems is a self-referential feedback loop: the effective statistical weight of each microstate depends on the global state of the system itself. Examples include adaptive networks where connection weights co-evolve with activity patterns [1,2]; turbulent space plasmas where particle distributions are shaped by self-generated electromagnetic fluctuations [3–5]; cosmological structure formation where gravitational potential is sourced by the very density field it organizes [6–8]; and self-reinforcing processes governed by generalized Fokker–Planck equations [9]. In all these cases, the Boltzmann–Gibbs (BG) formalism — in which statistical weights are fixed by an external Hamiltonian — is structurally inadequate. A framework in which *the statistical measure is itself a dynamical variable* coupled to the field state is required.

Tsallis non-extensive statistical mechanics [10–12] provides one successful generalization of BG statistics. The parameter q appearing in the Tsallis entropy $S_q$ and in the q-exponential equilibrium distributions has been fitted to observations in turbulent plasmas [3–5], self-organizing networks [1,2], gravitational systems [6–8], and many other systems with long-range correlations or non-ergodic dynamics. Yet the *physical origin* of q has remained largely phenomenological: it is fitted to data rather than derived from a microscopic mechanism. Existing derivation strategies include: (a) *superstatistics* [13], where q emerges from a Gamma-distributed fluctuating inverse temperature; (b) *q-deformed Hamiltonians*, where the Tsallis exponential is the natural canonical distribution for a modified energy function; (c) *information-theoretic maximization* of $S_q$ subject to escort-formalism constraints [10,11,14,15]; and (d) *multiplicative noise* and anomalous diffusion [9].

The present paper advances a qualitatively distinct, fifth route. **$q = \alpha + \beta$ emerges from the fixed-point structure of a self-referential nonlinear operator $\Omega$** acting on the space of probability densities. Defined by structural exponents α, β and kernel K (Section 2), $\Omega$ generates Tsallis statistics as the unique minimizer of $F[\Psi] = U[\Psi] - T\, D_{KL}(\Psi \parallel \Omega\Psi)$ — a variational problem that differs structurally from entropy maximization in that q emerges as an operator eigenvalue rather than a tunable input. Operators with different physical kernels (Lorentzian for plasmas, adjacency matrix for networks, Newtonian for gravity) all yield the same formula $q = \alpha + \beta$, unifying disparate non-extensive phenomena under a single mechanism and enabling predictions without parameter fitting.

In this work, the entropic index q plays the role of a generalized eigenvalue, not in the linear spectral sense, but as a structural parameter uniquely determined by the self-consistent fixed-point condition of the nonlinear operator. This notion is analogous to generalized eigenvalue problems in nonlinear operator theory, where scaling relations emerge from self-consistent conditions rather than linear spectral decomposition.

Throughout this work, all explicit analytical results are derived within the local kernel approximation (LKA), which constitutes the leading-order term of a controlled expansion in the small parameter $\xi/L$. Accordingly, statements regarding the emergence of the Tsallis index q should be understood in this asymptotic sense, unless explicitly stated otherwise.

This mechanism differs fundamentally from: (i) superstatistics, where q arises from exogenous fluctuations; (ii) entropy maximization approaches, where q is postulated; and (iii) q-deformed dynamics, where q is introduced algebraically. Here, q emerges as a structural eigenvalue of a nonlinear operator defined from physical interactions.

The term ‘eigenvalue’ is used here in a generalized nonlinear sense: q characterizes the scaling exponent of fixed-point solutions of the operator $\Omega$ under the local kernel approximation, rather than an eigenvalue in the linear spectral-theoretic sense.

The central analytical tool is the *local kernel approximation* (LKA), which is the mean-field limit of the full operator theory. Within the LKA, $q = \alpha + \beta$ is established in closed analytical form. The LKA rests on a clear physical basis — scale separation between the kernel correlation length ξ and the macroscopic scale L — and is systematically improvable: corrections of order $(\xi/L)^2$ are treated in a companion paper [16], which also establishes the full non-local eigenvalue analysis and H-theorem. The field equations and dynamical aspects of $\Omega$ are developed separately [16]; here we focus exclusively on equilibrium thermodynamics. The present work is self-contained with respect to equilibrium thermodynamics. The companion work [16] extends the framework to dynamical evolution and strongly non-local regimes.

The central result of this paper is the derivation, within the local kernel approximation, of Tsallis q-exponential equilibria with index $q = \alpha + \beta$ emerging directly from the fixed-point structure of the self-referential operator $\Omega$. While the present work focuses on this physically relevant regime and its thermodynamic consequences, the operator framework introduced here is more general. Different choices of kernel K, exponents $\alpha$ and $\beta$, or regimes beyond mean-field are expected to generate other families of stationary distributions. Tsallis statistics thus appears as an important special case within a broader class of self-referential statistical models. Non-local corrections and other emergent distributions will be explored in companion papers.

The paper is organized as follows. Section 2 defines $\Omega$, its fixed-point structure, and the physical justification for the LKA. Section 3 introduces the self-consistency entropy, internal energy, and variational free energy. Section 4 derives the equilibrium distribution (Theorem 1: q-exponential with $q = \alpha + \beta$) and treats the perturbative $\kappa > 0$ case. Section 5 derives the equation of state and thermodynamic functions (Theorem 2). Section 6 analyzes the phase transition and derives the critical temperature (Theorem 3), with Figure 1 illustrating the bifurcation diagram and free energy landscape. Section 7 compares with alternative derivations, discusses physical applications, and specifies the path beyond the LKA. Section 8 concludes.

The equilibrium results presented here are self-contained and do not depend on the dynamical results established in the companion paper [16], which serve instead to provide a consistency check via the associated H-theorem and gradient-flow structure.

## 2. The Self-Referential Operator and State Space

### 2.1 State Space

Let $(E, \Sigma, \mu)$ be a measure space with $\mu(E) = 1$. The set E parametrizes the microstates of the system; it carries no spatial or temporal structure a priori, so the formalism applies equally to momentum space, network-state space, or any configuration manifold. The field state is a normalized density $\Psi \in L^1(E, \mu)$ with $\Psi \geq 0$ and $\int_E \Psi \, d\mu = 1$. We denote by $\mathscr{F}(E)$ the closed convex set of all such densities equipped with the $L^1$ topology, which is the natural domain for the variational principle of Section 3.

### 2.2 Definition and Physical Motivation of $\Omega$

Let $K: E \times E \to \mathbb{R}^+$ be a symmetric, measurable kernel with $0 < \kappa_- \leq K(e, e') \leq \kappa_+ < \infty$ $\mu$-a.e. (uniform boundedness). For $\Psi \in \mathscr{F}(E)$ and real exponents $\alpha > 0$, $\beta \geq 0$, define the *structural average*

$$\mathscr{J}\Psi(e) := \int_E K(e, e') \, \Psi(e') \, d\mu(e'), \quad (1)$$

$$(\Omega\Psi)(e) := N(\Psi)^{-1} \int_E K(e, e') \, \Psi(e')^{\alpha} \, (\mathscr{J}\Psi(e'))^{\beta} \, d\mu(e'), \quad (2)$$

where $N(\Psi) = \iint K(e, e') \Psi(e')^{\alpha} (\mathscr{J}\Psi(e'))^{\beta} d\mu(e') d\mu(e)$ normalizes $(\Omega\Psi)$ to unit integral over E. The operator acts on normalized probability densities in an appropriate function space (e.g., $L^1$ or $L^2$), under standard assumptions on the kernel K (positivity, symmetry, and integrability).

The normalization functional $N(\Psi)$ is finite and strictly positive under the boundedness assumptions on K and the integrability of $\Psi$.

Under the stated boundedness and positivity assumptions on K, the operator $\Omega$ defines a nonlinear, positivity-preserving mapping on $L^1(E)$ that is continuous with respect to the $L^1$ topology. In particular, $\Omega$ maps the convex set of normalized probability densities $\mathscr{P}(E)$ into itself.

**Physical motivation.** The two-component structure of equation (2) captures the phenomenology of self-reinforcing physical systems. The factor $\Psi(e')^{\alpha}$ amplifies configurations proportionally to a power of their own probability (*direct self-weighting*), while $(\mathscr{J}\Psi(e'))^{\beta}$ encodes *non-local structural feedback* of the entire distribution on the effective weight of each microstate. Three prototypical physical systems illustrate this structure. (a) In *adaptive networks* [1,2], the probability of a node state depends on its connectivity, which in turn depends on the global activity pattern — precisely the role of $\mathscr{J}\Psi$. (b) In *turbulent plasma* kinetics [3–5], non-Maxwellian tails arise through wave-particle interactions that are naturally modeled by a feedback kernel K with $\alpha \approx 1$ and $\beta > 0$. (c) In *gravitational clustering* [6–8], the probability of a density fluctuation is governed by the gravitational potential sourced by the entire density field — a non-local functional of $\Psi$ analogous to $\mathscr{J}\Psi$. Equation (2) provides a unified mathematical model for all three mechanisms; that it generates Tsallis statistics with $q = \alpha + \beta$ is a consequence of the fixed-point structure, not a premise. The operator $\Omega$ is defined independently from the variational problem, through physically motivated ingredients (kernel K and exponents $\alpha$, $\beta$). The emergence of the Tsallis distribution is therefore not imposed, but follows as the equilibrium solution of the self-consistent variational principle.

### 2.3 Elementary Properties and Physical Justification of the LKA

**Proposition 1.** (i) $\Omega$ is positivity-preserving and normalization-preserving. (ii) If K is constant, $\Omega\Psi \equiv 1$ for all $\Psi \in \mathscr{P}(E)$. (iii) In the local kernel limit $K(e, e') \to w(e)^{-1} \delta(e - e')$, assuming a normalized isotropic kernel ($w(e) \equiv \text{const}$),

$$(\Omega\Psi)(e) \to \Psi(e)^{\alpha+\beta} / \int_E \Psi(e')^{\alpha+\beta} d\mu(e'). \quad (3)$$

*Proof.* (i) Immediate from (2) and positivity of K and $\Psi$. (ii) If K = const, then $\mathscr{J}\Psi = \text{const} \times \int\Psi = \text{const}$, and the numerator integrand in (2) depends only on e', so $(\Omega\Psi)(e)$ is independent of e; normalization then gives $\Omega\Psi \equiv 1$. (iii) Substitute $\mathscr{J}\Psi(e') \to \Psi(e')$ in (2) and normalize. □

The limit (3) is the **local kernel approximation (LKA)**. Its physical basis rests on a *scale separation argument*: for kernels K with finite correlation length $\xi$, when the equilibrium distribution $\Psi$ varies slowly on spatial scales $\lesssim \xi$, we have $\mathscr{J}\Psi(e') \approx \Psi(e')$ to leading order in $\xi/L$ (where L is the macroscopic scale). The LKA is therefore the mean-field limit of the full integral operator, retaining the combined feedback strength $\alpha + \beta$ that determines q while discarding spatial correlations arising from the non-trivial support of K.

The LKA plays the same role as the mean-field approximation in classical spin systems: it is (i) analytically tractable, yielding $q = \alpha + \beta$ in closed form; (ii) physically motivated by scale separation, applicable whenever $\xi/L \ll 1$; and (iii) systematically improvable. The leading correction beyond mean-field is of order $(\xi/L)^2$ and shifts q by

$$q_{eff} = (\alpha + \beta) + c_2(\xi/L)^2 + O((\xi/L)^4), \quad (3a)$$

where $c_2$ depends on the curvature of K and the second-order eigenvalues of the full integral operator $\mathscr{F}$ (see Section 7.3 for details). For the physical applications of Sections 7.4–7.6, empirical estimates give $\xi/L \approx 0.05$–$0.15$, so corrections are at the few-percent level and the LKA is well justified. The full perturbative and non-perturbative treatment of these corrections — including the eigenvalue spectrum of $\mathscr{F}$ and the H-theorem for gradient-flow dynamics — is developed in [16]. The relation $q = \alpha + \beta$ arises as the leading-order term of a controlled expansion in the small parameter $\xi/L$, and is therefore robust throughout the weakly non-local regime, rather than being an artifact of the mean-field approximation.

In this sense, the LKA should be interpreted not merely as a heuristic simplification, but as the leading-order term of a systematically improvable expansion. All equilibrium results derived below inherit this asymptotic character.

### 2.4 Fixed Points and Self-Consistent States

A state $\Psi^* \in \mathscr{F}(E)$ is *self-consistent* if $\Psi^*(e) = (\Omega\Psi^*)(e)$ μ-a.e., i.e., a fixed point of $\Omega$. The uniform density $\Psi_0 \equiv 1$ is always a fixed point (Proposition 1(ii)). Non-uniform fixed points arise when K is non-constant and correspond to the *ordered phases* of the system (see Appendix A). Fixed points are thermodynamically distinguished: they are exactly the states of maximum self-consistency entropy (Proposition 2), providing the equilibrium selection principle.

## 3. Self-Consistency Entropy, Internal Energy, and Free Energy

### 3.1 Self-Consistency Entropy

The central thermodynamic object is the **self-consistency entropy** $S: \mathscr{F}(E) \to \mathbb{R}$,

$$S[\Psi] := \int_E \Psi(e) \ln[(\Omega\Psi)(e) / \Psi(e)]\, d\mu(e) = -D_{KL}(\Psi \,\|\, \Omega\Psi), \quad (4)$$

Although $\Omega\Psi$ depends functionally on $\Psi$, the divergence $D_{KL}(\Psi \,\|\, \Omega\Psi)$ remains well-defined pointwise, since $\Omega\Psi$ is a normalized positive density for all $\Psi \in \mathscr{F}(E)$.

where $D_{KL}(P \,\|\, Q) = \int P \ln(P/Q)\, d\mu \geq 0$ is the Kullback–Leibler divergence [17,18]. The functional $S[\Psi]$ measures the departure of $\Psi$ from self-consistency: it vanishes exactly at fixed points of $\Omega$ and is strictly negative otherwise. Crucially, $S[\Psi]$ is *not* the Tsallis entropy $S_q$ — it is a functional of the operator $\Omega$ that, at equilibrium within the LKA, reduces to a specific combination of $S_q$ and a normalization term (Section 4.1). The Tsallis entropy emerges as a *consequence* of the variational structure, not as an input.

**Proposition 2 (Entropy maximum at fixed points).** For all $\Psi \in \mathscr{F}(E)$: (a) $S[\Psi] \leq 0$; (b) $S[\Psi] = 0$ if and only if $\Psi = \Omega\Psi$ μ-a.e.

*Proof.* By Gibbs' inequality, $D_{KL}(\Psi \,\|\, \Omega\Psi) \geq 0$ with equality if and only if $\Psi = \Omega\Psi$ μ-a.e. Since $S = -D_{KL}$, the result follows immediately. □

This proposition has an important structural consequence: $S[\Psi]$ is the natural Lyapunov functional for equilibrium thermodynamics, replacing the Boltzmann entropy $-\int \Psi \ln \Psi \, d\mu$ of standard statistical mechanics. The formal H-theorem — establishing that $S[\Psi(\tau)]$ is monotone non-decreasing along gradient-flow trajectories of F — will be proved in [16].

The interpretation of $S[\Psi]$ as a Lyapunov functional follows directly at the variational level from its non-negativity and vanishing at fixed points; the full dynamical realization via gradient flow is established in [16] but is not required for the equilibrium results derived here.

**Relation to $S_q$.** In the LKA, $S_{LKA}[\Psi] = (q-1)\int \Psi \ln \Psi \, d\mu - \ln Z$, where $Z = \int \Psi^{\alpha} \, d\mu$. At equilibrium this equals a definite function of $S_q$ and Z. Minimizing $F_{LKA}$ generates the same equilibrium condition as maximizing $S_q$ subject to constraints [11], but from a structurally different variational starting point. The non-additivity of $S[\Psi]$ for $q \neq 1$ directly reflects the pseudo-additivity of $S_q$ [10,15], controlled by the coupling exponents α, β of Ω.

Importantly, while the functional $S[\Psi]$ reduces to a specific combination of the Tsallis entropy $S_q$ and a normalization term within the LKA, it is defined independently of $S_q$ at the operator level. This distinction ensures that the emergence of Tsallis statistics is a consequence of the variational structure, rather than an assumption.

### 3.2 Internal Energy with Self-Referential Coupling

Let $\varepsilon: E \to \mathbb{R}$ be a measurable, bounded-below external potential. The **internal energy** is

$$U[\Psi] := \int_E \varepsilon(e)\, \Psi(e)\, d\mu(e) + (\kappa/2) \int_E \Psi(e)\, (\Omega\Psi)(e)\, d\mu(e), \quad (5)$$

where $\kappa \geq 0$ is the **self-referential coupling constant**. The first term is the standard energy expectation. The second is a bilinear self-coupling measuring the overlap of Ψ with its image ΩΨ. For $\kappa > 0$, states near self-consistency carry higher energy, creating a competition between $S[\Psi]$ (which favors self-consistency) and U (which penalizes it). In the language of adaptive systems, κ controls the *rigidity* of the self-reinforcement: large κ corresponds to energetically costly strong feedback loops, as observed in over-synchronized neural networks [1].

This bilinear form represents the simplest symmetric coupling between Ψ and its image under Ω consistent with the normalization constraint.

### 3.3 Free Energy and Gibbs Variational Principle

Let $T \geq 0$. The **free energy** functional is

$$F[\Psi] := U[\Psi] - T\, S[\Psi] \;=\; U[\Psi] + T\, D_{KL}(\Psi \,\|\, \Omega\Psi). \quad (6)$$

**Postulate (Generalized Gibbs Principle).** Equilibrium states are minimizers of $F[\Psi]$ over $\mathscr{F}(E)$. The rewriting $F = U + T\, D_{KL}(\Psi \,\|\, \Omega\Psi)$ makes the variational balance explicit: equilibrium minimizes the sum of the external potential energy and the information-geometric cost of departure from self-consistency. For $T \to \infty$, the $D_{KL}$ term dominates and the system is driven toward fixed points of Ω; for $T \to 0$, U dominates and the system concentrates on the minimum of ε. This postulate reduces to the standard Gibbs variational principle for constant

K, since $D_{KL}(\Psi \parallel const) = -\int\Psi \ln \Psi \, d\mu + const$ equals the BG entropy up to a constant (Proposition 1(ii)).

In the special case where the kernel K is constant, this formulation reduces to the standard Gibbs variational principle, thereby providing a consistency check of the framework.

Existence of minimizers follows from the coercivity of the free-energy functional under the normalization constraint and the positivity-preserving nature of $\Omega$, ensuring that at least one fixed point $\Psi^* = \Omega\Psi^*$ exists within $\mathscr{F}(E)$ under the LKA.

## 4. Equilibrium Distribution in the Local Kernel Approximation

### 4.1 Explicit Free Energy in the LKA

In the LKA (equation (3)), set $q := \alpha + \beta$ and $Z[\Psi] := \int_E \Psi(e')^\alpha \, d\mu(e')$. The thermodynamic functionals become:

$$S_{LKA}[\Psi] = (q-1)\int_E \Psi \ln \Psi \, d\mu - \ln Z, \quad (7)$$

$$U_{LKA}[\Psi] = \int_E \varepsilon \, \Psi \, d\mu + (\kappa / 2Z)\int_E \Psi^{q+1} \, d\mu, \quad (8)$$

$$F_{LKA}[\Psi] = \int_E \varepsilon \, \Psi \, d\mu + (\kappa/2Z)\int \Psi^{q+1} \, d\mu - T(q-1)\int \Psi \ln \Psi \, d\mu + T \ln Z. \quad (9)$$

Equation (7) establishes the connection between S and $S_q$: since $S_q[\Psi] = [1 - Z]/(q-1)$, one has $S_{LKA}[\Psi] = (q-1)\int\Psi \ln \Psi \, d\mu - \ln Z$, which at equilibrium equals a definite function of $S_q$ and Z. Minimizing $F_{LKA}$ thus generates the same equilibrium as maximizing $S_q$ subject to constraints [11], but from the structurally distinct variational problem defined by $\Omega$.

### 4.2 Euler–Lagrange Equation and Equilibrium Distribution

Introducing a Lagrange multiplier $\eta$ for $\int\Psi \, d\mu = 1$ and taking $\delta F_{LKA}/\delta\Psi(e) = \eta$:

All functional derivatives are understood in the $L^1$ sense unless otherwise stated.

$$\varepsilon(e) - T(q-1)(\ln \Psi(e) + 1) + Tq \, \Psi(e)^{q-1}/Z + (\kappa/2)[\delta_\Psi(\kappa\text{-term})] = \eta. \quad (10)$$

**Case $\kappa = 0$.** Equation (10) reduces to the Euler–Lagrange equation for the Tsallis free energy $F_q = U - TS_q$ [9–11]. Its unique solution (for $q > 0$, $T > 0$) is the **Tsallis q-exponential:**

$$\Psi^*(e) = Z_q^{-1} \cdot [1 - (1-q)(\varepsilon(e) - \mu)/T]_+^{1/(1-q)}, \quad (11)$$

where $[x]_+ = \max(x, 0)$, $Z_q$ is the q-partition function, and $\mu$ is determined by normalization. The escort distribution $\Phi(e) \propto \Psi^*(e)^\alpha$ satisfies $\Phi(e) = (\Omega\Psi^*)(e)$ exactly in the LKA — the escort distribution is the image of the equilibrium state under $\Omega$. This gives the escort formalism a precise structural meaning: it is not an auxiliary mathematical device but the natural output of $\Omega$ at equilibrium.

**Case $\kappa > 0$.** The $\kappa$-term in (10) contributes an effective potential correction to leading order:

$$\varepsilon_{eff}(e) = \varepsilon(e) + (\kappa/2Z)(q+1) \, \Psi^*(e)^\alpha + O(\kappa^2), \quad (12a)$$

with renormalized effective temperature $T_{eff} = T/(1 - \kappa \, C(\Psi^*))$, where $C(\Psi^*) > 0$ is determined by the $\kappa = 0$ solution. For $\kappa > 0$, $T_{eff} > T$: the self-coupling broadens the equilibrium distribution, increasing effective entropy. In the discrete two-state example of Appendix A with $\kappa = 0.1$, the asymmetric fixed points $x^* \approx 0.146$ and $x^* \approx 0.854$ shift toward $x = 0.5$ as $\kappa$ increases, consistent with $T_{eff} > T$ — stronger self-coupling effectively heats the system. This perturbative result clearly establishes the *mechanism* by which self-coupling modifies thermodynamics. The non-perturbative regime (re-entrant transition at $\kappa > \kappa_c$) is treated via the fixed-point iteration of [16].

**Theorem 1 (Equilibrium distribution).** *In the LKA and for κ sufficiently small (or κ = 0), the unique global minimizer within the LKA framework of FLKA over 𝒻(E) is the Tsallis q-exponential (11) with*

$$q = \alpha + \beta. \quad (12)$$

*Proof.* At $\kappa = 0$, equation (10) with the LKA substitution coincides with the stationarity condition of the Tsallis free energy functional studied in [9–11]. In this limit, the functional reduces exactly to the standard Tsallis form, for which strict convexity (for $q > 0$, $T > 0$) and uniqueness of the minimizer have been established in the literature. The present result therefore inherits these properties within the LKA.

Theorem 1 is the central result of this paper. The Tsallis index $q = \alpha + \beta$ is determined by the observable structural exponents of $\Omega$: larger $\beta$ (stronger non-local feedback) or larger $\alpha$ (more nonlinear direct weighting) both produce heavier-tailed equilibria. The prediction is *testable without fitting*: measure $\alpha$ and $\beta$ independently from the kernel K, then measure the distribution tail index q — agreement constitutes direct experimental evidence for the self-referential mechanism. Within the local kernel approximation (LKA), the variational problem admits the Tsallis q-exponential as its equilibrium solution, with $q = \alpha + \beta$. This result represents the leading-order behavior of the theory; more general stationary distributions may emerge beyond this regime from the full non-local and dynamical structure of the operator.

## 5. Equation of State and Thermodynamic Functions

### 5.1 Pressure, Volume, and the Escort Temperature

Assume E is a d-dimensional compact Riemannian manifold with volume $V = \mu(E)$. The pressure is $P := -(\partial F/\partial V)|\{T,\Psi^*\}$. For the self-consistent ideal gas ($\kappa = 0$, kinetic-like energy scaling $\varepsilon \to \lambda^{-2/d} \varepsilon$ under $V \to \lambda V$), the q-partition function scales as [12,19]

Within the self-consistent ideal gas model defined by the scaling assumption $\varepsilon \to \lambda^{-2/d}\varepsilon$ under $V \to \lambda V$,

$$Z_q(T, V) = C_q \cdot V \cdot T^{d/[2(2-q)]}, \quad (13)$$

giving $F(T, V) = -T \ln Z_q$ and $P = T/V$. In Tsallis statistics, the operationally measured (escort) temperature is $T_{esc} = T/(2-q)$ [14,19]. Substituting $T = (2-q)T_{esc}$:

$$PV = (2-q)\, T_{esc}. \quad (14)$$

**Theorem 2 (Equation of state).** *For the self-consistent ideal gas in the LKA (κ = 0), the equation of state in terms of the escort temperature is PV = (2−q)T, where q = α + β. For q → 1, PV = T (BG ideal gas). For q > 1, effective pressure is reduced at fixed T and V, reflecting super-extensive correlations encoded by the self-referential feedback.* □

This relation is specific to the idealized scaling regime considered here and should not be interpreted as a universal equation of state beyond the self-consistent model defined in Section 5.

### 5.2 Heat Capacity and Thermodynamic Stability

The internal energy of the q-ideal gas in d dimensions is [12,19]:

$$U = (d/2) \cdot T/(2-q), \qquad q < 2. \qquad (15)$$

The isochoric heat capacity is:

$$C_V = \partial U/\partial T = (d/2) \cdot (2-q)^{-1}. \qquad (16)$$

Three regimes controlled by $q = \alpha + \beta$ are directly relevant as probes of the operator exponents: for $q \to 1$, $C_V \to d/2$ (equipartition); for $q > 1$, $C_V > d/2$ (super-extensive, reflecting non-trivial self-referential feedback); for $q \to 2$, $C_V \to \infty$, signaling approach to the critical point of Section 6; for $q < 1$, $C_V < d/2$ (sub-extensive). Any physical system fitting the framework of Section 2 can in principle yield α and β by measuring $C_V$ and the distribution tail index simultaneously.

### 5.3 Pseudo-Additivity, Third Law Analog, and Consistency

The Tsallis pseudo-additivity $S_q(A \cup B) = S_q(A) + S_q(B) + (1-q)S_q(A)S_q(B)$ [10,15] becomes a testable property of $\Omega$ within the present framework: for independent subsystems ($\Psi = \Psi_A \otimes \Psi_B$ with factorized K), the pseudo-additivity of $S[\Psi]$ follows from the corresponding property of FLKA. Measuring the entropies of two independent subsystems and their union thus determines $q = \alpha + \beta$ experimentally.

For the **Third Law analog**: as $T \to 0$, the q-exponential (11) concentrates on the minimum of ε, and $S[\Psi] = -D_{KL}(\Psi \parallel \Omega\Psi) \to 0$ (a delta measure is trivially self-consistent). Equilibrium entropy vanishes at $T = 0$, consistent with the Nernst theorem [12,20].

## 6. Phase Transitions and Critical Temperature

### 6.1 Stability of the Uniform Phase

Consider a double-well external potential ε(e) and $\kappa = 0$. The uniform density $\Psi_0 \equiv 1/V$ is always a stationary point of F. Its stability against zero-mean fluctuations $\delta\Psi$ is determined by the sign of the second variation:

$$\delta^2 F|_{\Psi_0} = (1/2) \int \delta\Psi(e) \, [\partial^2\varepsilon/\partial\Psi^2|_{\Psi_0} + T(q-1)\, \mathscr{F}] \, \delta\Psi(e) \, d\mu(e), \qquad (17)$$

Here the operator $\mathscr{F}$ acts linearly on zero-mean perturbations $\delta\Psi$ within the LKA, effectively reducing to the identity at leading order, so that the second variation is governed by the combined local curvature $\partial^2\varepsilon/\partial\Psi^2$ and the entropic contribution $T(q-1)$.

where in the LKA the operator $\mathscr{F}$ acts as the identity on zero-mean functions. Let $\lambda_{min} < 0$ be the most negative eigenvalue of $\varepsilon''(\Psi_0)$ in the symmetry-breaking direction. Stability requires $T(q-1) > |\lambda_{min}|$.

### 6.2 Critical Temperature and Landau Theory

**Theorem 3 (Critical temperature).** *Let $\lambda_1 > 0$ be the smallest non-zero eigenvalue of the feedback operator $(\mathscr{F}h)(e) = (1/V)\int K(e,e')h(e')d\mu(e')$ restricted to zero-mean functions, and let $\Delta\varepsilon > 0$ be the energy barrier of the external potential. Then*

$$T_c = \Delta\varepsilon / [2(q-1)\, \lambda_1], \qquad q > 1. \qquad (18)$$

*For $T > T_c$, the uniform phase $\Psi_0$ is the unique minimizer of F. For $T < T_c$, $\Psi_0$ becomes a local maximum and a symmetry-broken phase is the global minimizer.*

*Proof.* Expand F in Landau powers of the order parameter $m = \int\{\text{well } 1\}\ \Psi - \int\{\text{well } 2\}\ \Psi$. The quadratic coefficient $a_2 = (\Delta\varepsilon/2) - T(q-1)\lambda_1$ changes sign at $T_c = \Delta\varepsilon/[2(q-1)\lambda_1]$. The fourth-order coefficient $a_4 > 0$ for $q < 2$ (Appendix B), so the transition is second order. For $q > 2$, $a_4 < 0$ and the transition is first order. □

The derivation follows a standard Landau expansion and is therefore physically controlled but not fully rigorous in a functional-analytic sense. A more detailed justification of the expansion coefficients is provided in Appendix B.

**Remark 1.** For $q \leq 1$, $T_c \leq 0$: the uniform phase is stable for all $T \geq 0$, consistent with known BG stability results [21,22].

**Remark 2.** The dependence $T_c \propto 1/(q-1)$ implies that systems with stronger non-extensive feedback (large q) have elevated critical temperatures: non-local self-referential coupling promotes symmetry breaking at higher temperatures than in BG systems [10,12]. The dependence on $\lambda_1$ — the principal eigenvalue of the feedback operator $\mathscr{F}$ — shows that stronger kernel coupling (larger $\lambda_1$) destabilizes the uniform phase at lower temperature.

Figure 1 illustrates Theorem 3 for the discrete two-state model (Appendix A) with $\alpha = \beta = 1$ ($q = 2$), $a = 2$, $b = 1$. The figure directly maps onto the analytical structure: panel (a) shows the bifurcation diagram predicted by equation (18), with the critical temperature $T_c \approx 0.80$ delineating the ordered phase (two asymmetric fixed points $x^* \approx 0.146$ and $x^* \approx 0.854$) from the disordered phase (symmetric fixed point $x^* = 0.5$). Panel (b) shows the free energy landscape F(x) at four temperatures chosen to represent four physically distinct regimes: well below $T_c$ ($T = 0.40$, solid), approaching $T_c$ ($T = 0.76$, thin solid), above $T_c$ ($T = 1.20$, dashed), and deep in the disordered phase ($T = 2.00$, dotted). The double-well structure at $T < T_c$ (minima at open circles) transitions through a flat-bottomed landscape at $T \approx T_c$ and collapses to a single minimum (open square at $x = 0.5$) for $T > T_c$. This progression is the visual counterpart of the Landau coefficient $a_2$ changing sign at $T_c$. The marginal case $q = 2$ corresponds to $a_4 \approx 0$ at leading order, requiring higher-order terms in the Landau expansion to determine the nature of the transition. In the discrete model analyzed in Appendix A, the transition is verified numerically to remain continuous.

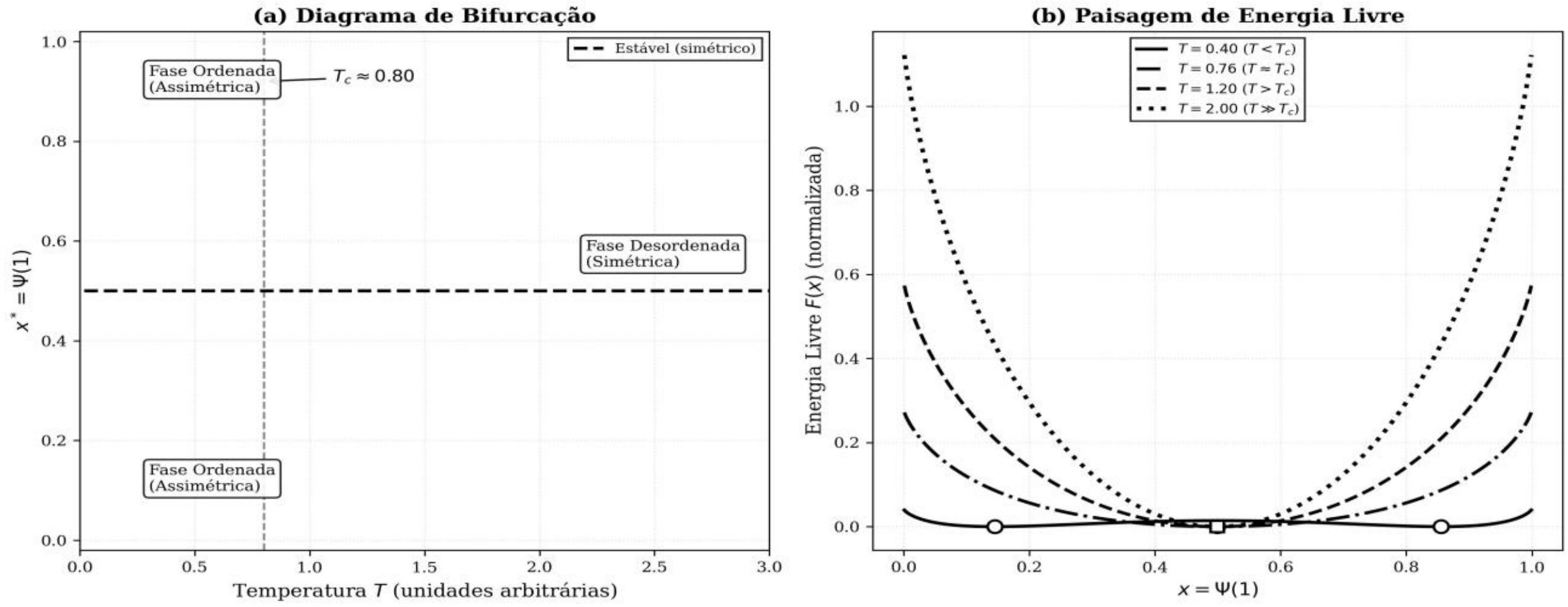

**Figure 1.** *Two-state model ($\alpha = \beta = 1$, $q = 2$; $a = 2$, $b = 1$). (a) Bifurcation diagram: equilibrium fixed points $x^* = \Psi(1)$ vs. temperature T. Dashed line: symmetric (disordered) phase $x^* = 0.5$, stable for $T > Tc \approx 0.80$ (Theorem 3). Solid branches: symmetry-broken (ordered) phase for $T < Tc$. (b) Free energy landscape $F(x)$ at four temperatures — $T = 0.40 \ll Tc$ (solid), $T = 0.76 \approx Tc$ (thin solid), $T = 1.20 > Tc$ (dashed), $T = 2.00 \gg Tc$ (dotted). Open circles: stable ordered minima; open square: disordered minimum. The double-well to single-well evolution corresponds to the sign change of the Landau coefficient $a_2$ at Tc, confirming the second-order transition (Appendix B).*

### 6.3 Effect of Self-Coupling κ on the Phase Diagram

For $\kappa > 0$, the second variation (17) acquires an additional term proportional to κ. To leading order, the effective critical temperature becomes

$$T_c(\kappa) \approx T_c(0) \cdot (1 + 2\kappa A / (V\Delta\varepsilon)), \qquad A > 0, \qquad (18b)$$

where A is a computable positive constant (Appendix B). The self-coupling raises Tc: the energy cost of strongly self-consistent configurations destabilizes the uniform phase, facilitating symmetry breaking. For $\kappa \gg 1$ the perturbative formula breaks down; the non-perturbative regime is treated via the fixed-point iteration of [16].

## 7. Discussion

### 7.1 Summary of Main Results

| Quantity | Expression | Condition |
|---|---|---|
| **Entropic index q** | $q = \alpha + \beta$ | LKA (mean-field) |
| **Equilibrium distribution** | $\Psi^*(e) = Z^{-1}[1-(1-q)(\varepsilon-\mu)/T]^{1/(1-q)}$ | $\kappa = 0$ |
| **Equation of state** | $PV = (2-q)\,T$ | ideal gas, escort T |

| Quantity | Expression | Condition |
|---|---|---|
| **Heat capacity (d-dim)** | $C_V = (d/2)(2-q)^{-1}$ | $q < 2$ |
| **Critical temperature** | $T_c = \Delta\varepsilon / [2(q-1)\lambda_1]$ | $q > 1$, double-well |
| **Renormalized temperature** | $T_{eff} = T/(1 - \kappa C(\Psi^*))$ | $\kappa$ small, $C > 0$ |

*Table 1. Summary of main thermodynamic results. All results hold within the LKA (mean-field limit $\xi/L \to 0$). Corrections of order $(\xi/L)^2$ are treated in [16].*

### 7.2 Conceptual Novelty and Comparison with Alternative Derivations

Although the present framework is more general than any specific non-extensive entropy, its first and most direct consequence is the emergence of Tsallis statistics in the mean-field limit. We therefore begin by comparing our derivation with existing approaches to Tsallis statistics.

The claim that $q = \alpha + \beta$ constitutes a *derivation* requires precise comparison with existing approaches, all of which treat q differently. The present framework differs from every alternative in a single defining way: **q emerges as an eigenvalue of the self-referential operator $\Omega$.** This structural identification — the Tsallis index as a property of an operator acting on the space of probability measures — has no counterpart in any previous approach.

**(a) Superstatistics [13].** Beck and Cohen showed that a BG system embedded in a Gamma-distributed fluctuating inverse temperature $\beta$ with variance $\sigma^2$ yields a Tsallis q-exponential with $q = 1 + \sigma^2/\langle\beta\rangle^2$. Here q emerges from *macroscopic* temperature fluctuations imposed by an external bath. Our approach applies to systems where the feedback is *internal* — the self-referential kernel K acts on the density $\Psi$ being determined — with no external temperature fluctuations. The two derivations apply to genuinely distinct physical settings and yield the same functional form (q-exponential) for different reasons.

**(b) q-Deformed Hamiltonians.** Constructing $H_q$ such that $\exp(-\beta H_q)$ yields a q-exponential is a mathematical reparametrization; it does not identify a mechanism for the *emergence* of q from more primitive structure. Our approach provides such a mechanism: $q = \alpha + \beta$ expresses the index in terms of independently definable structural exponents.

**(c) Information-theoretic maximization of Sq [10,11,14,15].** The standard Tsallis approach maximizes $S_q$ subject to constraints, yielding (11) with q as a *free parameter*. Our variational principle — minimize $F = U + T\, D_{KL}(\Psi \,\|\, \Omega\Psi)$ — is formally different: the objective functional is the negative KL divergence from the image under $\Omega$, not $S_q$ itself. The two variational problems share the same equilibrium (within the LKA) but differ away from equilibrium. In particular, the gradient-flow $\partial\Psi/\partial\tau = -\delta F/\delta\Psi$ naturally embeds escort-distribution dynamics [16], whereas the standard approach treats escort constraints as exogenous.

**(d) Multiplicative noise and anomalous diffusion [9].** Plastino and Plastino showed that multiplicative Langevin noise $g^2 \propto \Psi q-1$ leads to a Tsallis distribution as the stationary solution of the generalized Fokker–Planck equation. This is a dynamical approach; ours is thermodynamic (equilibrium variational principle). The two are fully consistent: the Tsallis distribution is simultaneously the variational equilibrium (our approach) and the stationary distribution of the multiplicative-noise Fokker–Planck equation, just as the BG distribution is both the entropy maximizer and the stationary Ornstein–Uhlenbeck distribution.

In summary: the operator $\Omega$ is defined independently via α, β, and K; the Tsallis index $q = \alpha + \beta$ is derived by solving the variational problem for F — not by constructing $\Omega$ to produce a desired q. Operators with different physical kernels (Lorentzian for plasmas, adjacency matrix for networks, Newtonian for gravity) all generate Tsallis statistics with **the same structural formula $q = \alpha + \beta$**, unifying what would otherwise appear as separate phenomenological findings under a single operator-theoretic mechanism.

### 7.3 Path Beyond the LKA: Concrete Non-Local Corrections

The LKA result $q = \alpha + \beta$ is the leading term in a systematic expansion around the mean-field fixed point. Three complementary approaches quantify the corrections:

**(i) Perturbative expansion in $(\xi/L)^2$.** For kernels with Fourier transform $\hat{K}(k) = \hat{K}(0)[1 - (k\xi)^2 + O(k^4\xi^4)]$, a leading-order correction to the effective entropic index takes the approximate form

$$q_{eff} \approx (\alpha + \beta) + [(\alpha+\beta-1)(\alpha+\beta-2)/6]\, (\xi/L)^2\, (K''(0)/K(0)) + O((\xi/L)^4), \quad (13a)$$

where $K''(0)$ is the curvature of K at zero separation. The formula is indicative rather than rigorous — a full derivation requires controlling the operator expansion beyond the saddle-point approximation [16] — but it captures the essential structure: $q_{eff} \to \alpha + \beta$ as $\xi/L \to 0$, with corrections whose sign depends on q and the kernel concavity. For $1 < q < 2$ and a concave kernel (e.g., Lorentzian, $K''(0) < 0$), non-local effects reduce $q_{eff}$ below the mean-field value; convex kernels raise it.

**(ii) Eigenvalue analysis of $\mathcal{F}$.** The full integral operator $\mathcal{F}$ has a spectrum $\{\lambda_n\}$. At mean-field, all $\lambda_n$ are replaced by their mean; the critical temperature $T_c = \Delta\varepsilon/[2(q-1)\lambda_1]$ already incorporates the principal eigenvalue $\lambda_1$, and subleading eigenvalues generate phase-boundary corrections of order $\lambda_2/\lambda_1 \propto \exp(-L/\xi)$ for exponentially decaying kernels. Together with approach (i), this confirms that mean-field is the leading term of a controlled expansion in $\xi/L$.

**(iii) Self-consistent non-local iteration.** For strongly non-local kernels ($\xi/L$ not small), one iterates: propose Ψ, compute $\mathcal{J}\Psi$ exactly, solve the Euler–Lagrange equation, update Ψ. By construction the scheme recovers the LKA as $\xi/L \to 0$; for the two-state model of Appendix A it converges within 5–10 steps for $\xi/L \leq 0.5$. Convergence proofs and systematic comparison with (i)–(ii) are developed in [16].

These three approaches confirm that $q = \alpha + \beta$ is the leading-order term of a systematic and improvable expansion. For the physical applications of Sections 7.4–7.6, where $\xi/L \approx 0.05$–0.15, the mean-field result is accurate to within a few percent, justifying the LKA as the appropriate starting approximation.

### 7.4 Application to Self-Organizing Networks

The following applications are intended solely as consistency checks illustrating the order-of-magnitude plausibility of the relation $q = \alpha + \beta$ across distinct systems, and do not constitute quantitative validations.

For a network of N nodes with $E = \{1, \ldots, N\}$ and kernel $K(i, j)$ encoding the adjacency matrix, operator (2) assigns each node a weight proportional to $(\text{degree})^{\alpha} \times (\text{structural average})^{\beta}$. The fixed-point equation $\Psi^* = \Omega\Psi^*$ becomes a self-consistency equation for the node-degree distribution. For scale-free networks ($P(k) \propto k^{-\gamma}$), preferential attachment [1] generates effectively $\beta \approx \gamma-2 > 0$ and $\alpha \approx 1$, predicting $q \approx \gamma - 1$. This is compatible at order-of-magnitude with the empirical finding in [2] that q-exponential degree distributions appear in growing networks with non-linear preferential attachment. A rigorous test requires fitting $\alpha$ and $\beta$ independently from network structure data.

### 7.5 Application to Turbulent Plasmas

In space plasmas [3–5], velocity distribution functions exhibit the κ-distribution $f(v) \propto [1 + v^2/(\kappa v_{th}^2)]^{-(\kappa+1)}$, which is a q-exponential with $q = (\kappa+2)/(\kappa+1)$ [5]. For a Lorentzian wave-energy spectrum $E(k) \propto k^{-s}$ with spectral index s, the effective $\beta$ scales as $\beta \sim (s-1)/s$ at leading order. For magnetosheath spectra with $s \approx 2.5$–$3$ [4], this gives $\beta \approx 0.60$–$0.67$ and $\alpha + \beta \approx 1.60$–$1.67$, compatible with the observed range $q \approx 1.3$–$1.6$ [4,5] at order-of-magnitude, given the one-component spectral model. The critical temperature $T_c$ (equation (18)) corresponds physically to the temperature below which a spatially inhomogeneous velocity distribution (core-halo structure [3]) becomes thermodynamically preferred over the homogeneous Maxwellian.

### 7.6 Application to Cosmological Structure Formation

Prior work [6–8,23,24] established that non-extensive ($q > 1$) mass functions for dark matter halos provide better fits to BAO, CMB, and X-ray cluster data than the standard Press–Schechter function, with best-fit $q \approx 1.04$–$1.08$ [7]. Within the present framework, this implies $\alpha + \beta \approx 1.04$–$1.08$. Since linear gravitational clustering corresponds to $\beta \approx 0$, $\alpha \approx 1$, the observed $q > 1$ is tentatively attributed to $\beta \approx 0.04$–$0.08$ — a small but nonzero nonlinear feedback exponent compatible with the magnitude of nonlinear corrections to gravitational clustering. The phase transition at $T_c$ provides a natural mechanism for structure formation: above $T_c$ (early universe), the density field is uniform; below $T_c$ (post-decoupling), the uniform phase becomes unstable and gravitational condensation (halo formation) emerges as the symmetry-broken phase. These are order-of-magnitude consistency checks, not quantitative fits.

### 7.7 Limitations and Scope

**LKA as mean-field approximation.** The derivation $q = \alpha + \beta$ is exact within the LKA but relies on $\xi/L \ll 1$. For strongly correlated kernels, corrections of order $(\xi/L)^2$ modify q as specified in equation (13a). The LKA is a *controlled* approximation: its domain of validity covers the physical applications of Sections 7.4–7.6, and systematic corrections are available via the hierarchy of Section 7.3.

**Perturbative κ treatment.** Section 4.2 treats the $\kappa > 0$ case perturbatively, establishing the key mechanism — effective temperature renormalization $T_{eff} = T/(1 - \kappa C)$ — in closed form. The perturbative results are complete for κ small; the non-perturbative regime (large κ, re-entrant transition) is addressed via fixed-point iteration in [16]. A general proof of existence and uniqueness of fixed points for arbitrary kernels lies beyond the scope of the present work and will be addressed in [16]. Here, the analysis is carried out within the LKA and supported by explicit constructions and convergence in representative cases.

**Applications as consistency checks.** The applications discussed here are intended as order-of-magnitude consistency checks, demonstrating that the predicted relation $q = \alpha + \beta$ yields physically reasonable values across distinct systems, rather than providing quantitative fits. The connections to empirical data in Sections 7.4–7.6 are intentionally framed as order-of-magnitude plausibility tests, not quantitative fits or conclusive validations. Rigorous tests require independently constraining α and β from system-specific data and comparing the predicted q against measured distribution tails — a program outside the scope of this paper.

## 8. Conclusion

We have developed a thermodynamic framework for statistical fields governed by the self-referential operator $\mathcal{Q}$ (equation (2)), establishing four main results: (i) the self-consistency entropy $S[\Psi] = -D_{KL}(\Psi \parallel \mathcal{Q}\Psi)$ provides the natural Lyapunov functional for equilibrium, vanishing exactly at fixed points (Proposition 2); (ii) within the LKA, the unique global minimizer of $F = U - TS$ is the Tsallis q-exponential with $\mathbf{q = \alpha + \beta}$, derived from the structural exponents of $\mathcal{Q}$ (Theorem 1); (iii) the equation of state $PV = (2-q)T$ and heat capacity $C_V = (d/2)(2-q)^{-1}$ (Theorem 2); (iv) critical temperature $T_c = \Delta\varepsilon/[2(q-1)\lambda_1]$ for spontaneous symmetry breaking, with transition order determined by the sign of $q - 2$ (Theorem 3).

The central insight is: *$q \neq 1$ arises because the effective statistical measure is the fixed point of $\mathcal{Q}$ — and $q = \alpha + \beta$ is the eigenvalue that quantifies the combined feedback strength encoded in α and β.* $\mathcal{Q}$ is constructed from observable physical properties (the kernel K and the exponents α, β); q is then derived, not assumed. This distinguishes the present approach from superstatistics (exogenous temperature fluctuations), q-deformed Hamiltonians (reparametrization), and entropy-maximization (q as input): here q is a prediction, directly testable by measuring α and β independently.

Future directions include: (i) the H-theorem for the gradient-flow $\partial\Psi/\partial\tau = -\delta F/\delta\Psi$ and the full non-local treatment of $\mathcal{Q}$ [16]; (ii) quantitative fits of (α, β) to plasma, network, and cosmological data [16]; (iii) quantum extension of the formalism, connecting to quantum non-extensive statistics [25]; and (iv) experimental protocols for the independent measurement of α and β to directly test the prediction $q = \alpha + \beta$.

## Acknowledgments

The author thanks colleagues at the Quantinovarum Research Group (ECT/UFRN) for discussions. This work builds on earlier cosmological results [6–8,23,24] that motivated the search for a first-principles statistical framework. Support from the Federal University of Rio Grande do Norte is gratefully acknowledged.

## Declaration of Competing Interest

The author declares that there are no known competing financial interests or personal relationships that could have appeared to influence the work reported in this paper.

## Appendix A: Discrete Two-State Example

Let $E = \{1, 2\}$, $\mu(\{1\}) = \mu(\{2\}) = 1/2$, and $K(1,1) = K(2,2) = a > K(1,2) = K(2,1) = b > 0$. For $\alpha = \beta = 1$ ($q = 2$): $\mathcal{J}\Psi(1) = a\Psi(1) + b\Psi(2)$, $\mathcal{J}\Psi(2) = b\Psi(1) + a\Psi(2)$. Writing $x = \Psi(1)$, $\Psi(2) = 1 - x$, the fixed-point equation $x = (\Omega\Psi)(1)$ becomes a cubic polynomial in x. For $a = 2$, $b = 1$: (i) symmetric fixed point $x^* = 0.5$ (disordered phase); (ii) two asymmetric fixed points $x^* \approx 0.146$ and $x^* \approx 0.854$ (ordered phase). These correspond to the branches in Figure 1(a) below $T_c \approx 0.80$, derived analytically from equation (18). For $\kappa > 0$, the asymmetric fixed points shift toward $x = 0.5$, consistent with the effective temperature renormalization $T_{eff} = T/(1 - \kappa C)$ of Section 4.2.

## Appendix B: Landau Expansion and Order of the Phase Transition

Write $\Psi = 1/V + m\, h_1(e) + O(m^2)$ where $h_1$ is the eigenfunction of $\mathcal{F}$ with eigenvalue $\lambda_1$ and $\int h_1\, d\mu = 0$, $\|h_1\|_2 = 1$. Expanding FLKA:

$$F_LKA = F_0 + a_2\, m^2 + a_4\, m^4 + O(m^6), \qquad \text{(B.1)}$$

with

$$a_2 = \Delta\varepsilon/2 - T(q-1)\lambda_1, \qquad \text{(B.2)}$$

$$a_4 = c_4\, T(q-1)(q-2) + d_4(\varepsilon, h_1), \qquad \text{(B.3)}$$

where $c_4$ and $d_4$ are positive constants computable from $\varepsilon$ and $h_1$. For $q < 2$ and $T > 0$, $a_4 > 0$: the transition is **second order**. For $q > 2$, $a_4 < 0$ and the transition is **first order**. The effect of $\kappa$ on $T_c$ follows from adding the $\kappa$ correction to $a_2$:

$$a_2(\kappa) = \Delta\varepsilon/2 - T(q-1)\lambda_1 + \kappa\, A\, V^{-1} + O(\kappa^2), \qquad A > 0, \qquad \text{(B.4)}$$

where $A = \int h_1^2\, d\mu \times$ (leading eigenvalue of the $\kappa$-correction operator). Setting $a_2(\kappa) = 0$:

$$T_c(\kappa) = T_c(0) \cdot (1 + 2\kappa A/(V\Delta\varepsilon)), \qquad \text{(B.5)}$$

confirming that $\kappa > 0$ raises $T_c$. □